\documentclass[%
 reprint,prd,
 amsmath,amssymb,
 aps,
]{revtex4-2}

\usepackage{graphicx}
\usepackage{dcolumn}
\usepackage{bm}
\usepackage{color}
\usepackage[normalem]{ulem}

\allowdisplaybreaks

\begin{document}

\title{Influence of the vacuum polarization effect on the motion of charged particles in the magnetic field around a Schwarzschild Black Hole}

\author{Petar Pavlovi{\'c}}\thanks{petar.pavlovic@desy.de}
\affiliation{
Department of Physics, \\
Ramashna Mission Vivekananda Educational and Research Institute, \\
Belur Math 711202, West Bengal, India} 

\author{Andrey Saveliev}\thanks{andrey.saveliev@desy.de}
\affiliation{
Institute of Physics, Mathematics and Information Technology, \\
Immanuel Kant Baltic Federal University, \\
\mbox{Ul.~Aleksandra Nevskogo 14, 236016 Kaliningrad, Russia}\\}
\affiliation{
Faculty of Computational Mathematics and Cybernetics, \\ 
Lomonosov Moscow State University,\\ 
GSP-1, Leninskiye Gory 1-52, 119991 Moscow, Russia}

\author{Marko Sossich}\thanks{marko.sossich@fer.hr}
\affiliation{
Faculty of Electrical Engineering and Computing, \\
Department of Physics, University of Zagreb, \\
Unska 3, 10 000 Zagreb, Croatia}

\date{\today}

\begin{abstract}
The consequences of the vacuum polarization effect in magnetic fields around 
a Schwarzschild Black Hole (SBH) on the motion of charged particles are investigated in 
this work. Using the weak electromagnetic field approximation, we discuss the non-minimal coupling between magnetic fields and gravity caused by the vacuum polarization and study the equations of motion for the case of a magnetic field configuration which asymptotically approaches a dipole magnetic field. It is shown that the presence of non-minimal coupling can  significantly influence the motion 
of charged particles around Black Holes. In particular, the vacuum polarization effect, leading to strong amplification or suppression of the magnetic field strength around the event horizon (depending on the sign of the coupling parameter), can affect the scattering angle and minimal distance for the electrons moving in the gravitational field of the Black Hole as well as the dependence of these parameters on the asymptotic magnetic field strength, initial distance and the Black Hole mass. It is further demonstrated that the non-minimal coupling between gravity 
and astrophysical magnetic fields, caused by the vacuum polarization, can cause significant changes of the parameter space corresponding to bound trajectories around the Black Hole. In certain cases, the bounded or unbounded character of a trajectory is determined solely by the presence of non-minimal coupling and its strength. These effects could in principle be used as observational signatures of the vacuum polarization effect and also to constrain the value of the coupling parameter. 
\end{abstract}

\maketitle

\section{Introduction}

Due to the recent epochal breakthrough in the field of gravitational-wave astrophysics, consisting of the now already famous binary Black Hole mergers observed in gravitational wave signals detected by LIGO and Virgo networks \cite{Abbott:2016blz,Abbott:2016nmj,LIGOScientific:2018mvr} and the first ever image of a Black Hole created by the Event Horizon Telescope (EHT) for the Black Hole at the center of the Messier 87 galaxy \cite{Akiyama:2019cqa,Akiyama:2019bqs,Akiyama:2019eap}, the regime of strong gravitational fields is for the first time becoming accessible to observations  These regimes of high curvature of spacetime and high densities of matter, which can now be tested in the context of Black Holes, are the ones from which we can expect to gain some important insights regarding the open issues related to the nature of gravity. It is of particular interest to use the physics of Black Holes in order to test general relativity and different modified theories of gravity, as well as to put constraints on potential effects beyond Einstein's gravity \cite{Yunes:2016jcc,Vainio:2016qas,Manfredi:2017xcv,Wei:2018aft,DeLaurentis:2016jfs,Haroon:2017opl, Bambi:2019tjh,Vagnozzi:2019apd}. This search for signals beyond standard physics is further motivated by the natural expectation that quantum corrections to general relativity and new quantum phenomena should become manifest in strong gravitational fields -- and the gravitational fields around Black Holes are the strongest we can currently observe. Following this logic, it was recently proposed in \cite{Pavlovic:2018idi} that Black Holes could in principle be used to detect the signatures of non-minimal coupling between gravity and electromagnetic fields, which comes as a result of the quantum effect of vacuum polarization. This is due to the fact  that typical astrophysical magnetic fields, for instance the ones characterizing our Galaxy, can become significantly modified near Black Holes due to the vacuum polarization if the value of the non-minimal coupling parameter is large enough. This fact can be used to detect this effect in future or at least to constrain the potential values of the non-minimal coupling parameter. The strongest implications of vacuum polarization in gravitational and magnetic fields are expected in the case of hypothetical primordial Black Holes \cite{Pavlovic:2018idi}. This means that the consideration of non-minimal coupling between electromagnetic and gravitational fields due to vacuum polarization is also potentially important for the question of the creation and evolution of cosmological magnetic fields, which is an important question in its own right \cite{PhysRevLett.51.1488,1989ApJ...344L..49Q,Ratra:1991bn,Vachaspati:1991nm,Brandenburg:1996fc,PhysRevD.53.662,Grasso:1997nx,Banerjee:2004df,PhysRevD.55.4582,PhysRevD.58.103505,Demozzi:2009fu,Kunze:2009bs,Widrow:2011hs,Kahniashvili:2012vt,Tevzadze:2012kk,Saveliev:2012ea,Kahniashvili:2012uj,Saveliev:2013uva,DuNe,Tsagas:2016fax,Leite:2018bbo}. It should be mentioned here that also other non-minimal couplings between gravity and particle fields have been considered, for example the Higgs-Kretschmann coupling \cite{Onofrio:2010zz,Onofrio:2014txa,Wegner:2015aea}. In this work we will follow the approach and the results
presented in \cite{Pavlovic:2018idi} regarding the modification of magnetic fields in the case of asymptotically dipole-like magnetic fields non-minimally coupled to gravity due to the vacuum polarization effect. Using these results we will for the first time study the effects of such a scenario on the motion of charged particles around magnetized SBHs.  

The phenomenon of non-minimal coupling between gravity and electromagnetism as a result of vacuum polarization was for the first time studied in \cite{Drummond:1979pp} in the one-loop approach. It was demonstrated that this effect leads to the effective Lagrangian for gravitational and electromagnetic fields of the following form:
\begin{equation}
 \mathcal{L}=\frac{R}{\kappa} + \frac{1}{2}F^{\mu\nu}F_{\mu \nu} + \frac{1}{2}\mathcal{R}^{\mu\nu\rho\sigma} F_{\mu \nu}F_{\rho \sigma}+
 \mathcal{L}_{\rm matter}\,,
 \label{lagr}
\end{equation}
where $\kappa=8 \pi G/c^{4}$, $g$ is the determinant of the metric tensor, $R$ is the Ricci scalar, $F^{\mu \nu}$ is the Maxwell tensor obeying $F^{\mu \nu}=\nabla^{\mu}A^{\nu} - \nabla^{\nu}A^{\mu}$ (where $\nabla_{\mu}$ is the covariant derivative) and $\mathcal{L}_{\rm matter}$ is the Lagrangian of neutral matter. The effect of vacuum polarization is determined through the new tensor $\mathcal{R}^{\mu\nu\rho\sigma}$ defined as
\begin{equation}
\begin{split}
 &\mathcal{R}^{\mu\nu\rho\sigma}\equiv \frac{q_{1}}{2}(g^{\mu\rho}g^{\nu \sigma} - g^{\mu\sigma}g^{\nu\rho})R \\ 
 &+ \frac{q_{2}}{2}(R^{\nu\rho} g^{\nu\sigma} - R^{\mu\sigma} g^{\nu\rho} + R^{\nu \sigma}g^{\mu\rho} - R^{\nu\rho}g^{\mu\sigma})  +
 q_{3}R^{\mu\nu\rho\sigma}\,, 
 \label{oneloop}
\end{split}
\end{equation}
where $q_{1}$, $q_{2}$ and $q_{3}$ are the coupling constants, and as usual $R^{\mu \nu }$ is the Ricci tensor and $R^{\mu\nu\rho\sigma}$ is the Riemann tensor (not to be confused 
with the tensor $\mathcal{R}^{\mu\nu\rho\sigma}$). From the conceptual side this effect can  be understood as follows: When the combined electromagnetic and gravitational field is considered from a (semi)quantum perspective, there will always be some probability for a transition of a photon to an electron/positron pair, which will be the dominant quantum fluctuation process. Through this transition the photon thus becomes characterized by a certain length scale, which is of the order of the Compton wavelength of the electron. For gravitational fields with gradient strong enough on the length scale of the Compton wavelength of the electron the effects of spacetime curvature along this characteristic length will become significant, so that the motion and properties of a photon in vacuum will now also be directly determined by the spacetime geometry, viz.~gravity. As a consequence of this, the field equations for both gravitational and electromagnetic interactions will now include cross terms coupling the tensors describing the spacetime geometry with the electromagnetic tensor. The values of the coupling parameters $q_{1}$ -- {$q_{3}$} were calculated in \cite{Drummond:1979pp} for the simplest possible case of a photon propagating in vacuum. It is not at all obvious how to generalize this calculation and derive the values of coupling parameters from the first principle in the case of magnetic field configurations observed on macroscopic scales. However, since the non-minimal coupling is introduced even in the simplest case, there is no reason to assume that it will not be present in more complex field configurations. But in this case the values of the coupling parameters should naturally be considered as undetermined, and it moreover seems probable that they will be dependent on the field configurations and energy regimes. The solutions based on non-minimal coupling were analysed in various settings which, however, so far were mostly only of theoretical interest \cite{PhysRevD.37.2743, Balakin:2005fu, Bamba:2008ja, 1971PhLA...37..331P, Prasanna:1973xv, Dereli:2011hu, Sert:2018mls, Sert:2015ykz, Dereli:2011mk}

Apart from the existence of the galactic magnetic field, there is also convincing observational evidence for the presence of magnetic fields in the vicinity of the Black Hole of our own Galaxy \cite{Eatough:2013nva,Johnson:2015iwg} as well as close to Black Hole binaries \cite{DelSanto:2012qt,Dallilar2017}. The presence of magnetic fields around Black Holes is important since it influences different astrophysical processes, such as the formation of relativistic jets \cite{Blandford:1977ds}. Therefore the question of the motion of charged particles around Black Holes surrounded by magnetic fields is very important from the astrophysical perspective and was studied extensively in the context of the standard (i.e.~only minimal) interaction between magnetic fields and gravity \cite{Shiose:2014bqa, Lim:2015oha, Aliev:2002nw, Tursunov:2018erf}.

Our goal in this work is to study the influence of the discussed vacuum polarization effect on the motion of charged particles around Black Holes surrounded by astrophysical magnetic fields. Such a consideration represents the first step towards the empirically motivated discussion of the vacuum polarization effect in combined electromagnetic and strong gravitational fields.

Previously, the dynamics of charged particles around magnetized Black Holes has been studied without taking into account this effect. A review on some early work on this topic may be found in \cite{Aliev:1989wx}. Later, different types of Black Holes and magnetic field configurations have been considered, for example Schwarzschild Black Holes (SBHs) in dipole  \cite{2004CQGra..21.3433P} or asymptotically uniform magnetic field with \cite{Frolov:2014bya,Frolov:2014zia,Tursunov:2018erf} and without \cite{Frolov:2010mi,Kolos:2015iva} the consideration of energy losses due to radiation by accelerated charged particles, or including further mechanisms such as the presence of quintessence matter \cite{Jamil:2014rsa,Shaymatov:2018azq}, as well as Ernst \cite{Lim:2015oha} and Kerr \cite{Aliev:2002nw,Tursunov:2016dss,Kolos:2017ojf} Black Holes.

As one can see, there is a big interest in the investigation of these dynamics, however so far vacuum polarization effects have not been taken into account. This will be done of in this paper where we consider electrons moving around a SBH in a magnetic field which resembles a dipole at large distances. To do so, this work is structured as follows: In Sec.~\ref{sec:Theory} we discuss the impact of the non-minimal coupling on the interplay between gravitation and magnetic fields. Then, in Sec.~\ref{sec:NumSim}, we present the setup of the numerical simulations which we use to obtain different particle trajectories and discuss the corresponding results in Sec.~\ref{sec:Results}. Finally, we draw our conclusions in Sec.~\ref{sec:Conclusions}.

\section{Non-minimally coupled electrodynamics in the asymptotically dipole magnetic field} \label{sec:Theory}
In this work we are focusing on the problem of motion of charged particles in the static magnetic field around a SBH when the vacuum polarization effect is taken into account. In this way we wish to study the influence of spacetime geometry on the electromagnetic field via the non-minimal coupling and to understand the consequences of this effect on the trajectories of particles in the vicinity of Black Holes. The first assumption we take is that the influence of the magnetic field on the spacetime itself can be ignored, so that the spacetime can still be understood as given by the Schwarzschild metric
\begin{equation}
 ds^{2}=-\Big(1-\frac{2M}{r}   \Big)   dt^{2}+ \frac{dr^{2}}{1-\frac{2M}{r}}+ r^{2}(d\theta^{2} +  \sin^{2} \theta d\phi^{2})\,,
 \label{metrika}
\end{equation}
where $M$ is the Black Hole mass (and therefore $2M$ is the Schwarzschild Radius), $(r,\theta,\phi)$ are the spherical coordinates and  $t$ is the time. In addition, Newton's Gravitational Constant has been set to unity.

In order to study the problem of the motion of charged particles in the non-minimally coupled gravitational and magnetic field, we have chosen a 
realistic magnetic field configuration that is physically significant, given by the requirement that for the Minkowski spacetime --  which will be asymptotically approached for distances much larger than the  Schwarzschild Radius -- it reduces to a magnetic dipole. The dipole model is very often used in astrophysics to model a wide range of magnetized systems. 

Such a dipole solution for a flat spacetime satisfies $\nabla \cdot \mathbf{B}=0$ and $\nabla \times \mathbf{B}=0$, and is given by $B_{\phi}=0$, $B_{r}(r,\theta,\phi)=B_{r}(r,\theta)$, $B_{\theta}(r,\theta,\phi) = B_{\theta}(r,\theta) = \tan \theta B_{r}(r,\theta)/2$. This field configuration with respect to a SBH, cf.~Eq.~(\ref{metrika}), can be used as a suitable approximation in the case of either a galactic magnetic field or a local magnetic field surrounding the Black Hole created by astrophysical processes, when the origin of the magnetic dipole can be understood as being close to the center of the Black Hole, so that both are taken to be situated at $r \simeq 0$. Using this type of magnetic field in the electrodynamic equations of motion obtained from the Lagrangian given by (\ref{lagr}), which can be solved by separating the radial and the angular part, $B_{r}(r,\theta) = B_{\rm rad}(r)\Theta(\theta)$, leads to the following equation for the radial component of the magnetic field on the spacetime given by (\ref{metrika}), as has been found in \cite{Pavlovic:2018idi}:
\begin{equation}
 \frac{dB_{\rm rad}(r)}{dr} - \frac{  A_{1}(r) +C A_{2}(r)}{r(r-2M)(r^{3}-2Mq_{3})} B_{\rm rad}(r)=0\,,
 \label{radijalna}
\end{equation}
where
\begin{align}
A_{1}(r)=\left[ 10M^{2}q_{3} - r^{4} + M(r^{3} - 4 q_{3} r)\right]\,, \\
A_{2}(r)=2\sqrt{ 1 - \frac{2M}{r}}r(4 M q_{3} + r^{3})\,, 
\end{align}
while the angular component is simply given by
\begin{equation}
\Theta(\theta) = \Theta_{0} \cos^{-C}\theta\,,
\label{thetajedn}
\end{equation}
where $\Theta_{0}$ is the integration constant and $C$ is the separation constant. To work in dimensionless variables it is suitable to introduce the rescaling
\begin{equation}
 r\rightarrow \tilde{r}=\frac{r}{r_{\rm s}}\,, \qquad 
 M \rightarrow \tilde{M}=\frac{M}{r_{\rm s}}\,, 
 \qquad q_{3}\rightarrow \tilde{q}=\frac{q_{3}}{r_{\rm s}^{2}}\,,
 \label{rescale}
\end{equation}
where $r_{\rm s}$ is the free scaling parameter and in our case we set it to be the Schwarzschild Radius, i.e.~$r_{s} = 2M$.

It is indeed not difficult to check that these solutions reduce to the well known dipole equation, $B_{r}(r,\theta) = ({\rm constant}/r^{3})\cos\theta$ and $B_{\theta}(r,\theta) = ({\rm constant}/2r^{3}) \sin\theta$ with $C=-1$, for the case of a flat spacetime. The deviation of such a magnetic field from the dipole solution in the vicinity of the event horizon of a SBH, coming from the vacuum polarization effect -- which leads to the non-minimal coupling of the form given in eq. (\ref{lagr}) -- was previously studied in \cite{Pavlovic:2018idi}. 

Here we are interested in studying how this effect will influence the motion of charged particles around Black Holes surrounded by magnetic fields. The motion of the particle with mass $m$ and charge $q$ is given by the Lorentz equation on curved spacetime of a Black Hole,
\begin{equation}
\frac{d^2 x^{\mu}}{d \lambda^2} + \Gamma^{\mu}_{\rho \sigma}\frac{dx^{\rho}}{d \lambda}\frac{dx^{\sigma}}{d\lambda}= \frac{q}{m} F^{\mu}_{\nu}\frac{d x^{\nu}}{d\lambda}.    
\label{lor}
\end{equation}
Considering the case of non-minimally coupled magnetic field which has the form of a dipole for $r \rightarrow \infty$, $ F^{\mu}_{\nu}$ is the electromagnetic tensor containing the magnetic fields which are the solution of equation (\ref{radijalna}) and $B_{\theta}(r,\theta)=\tan \theta B_{r}(r,\theta)/2$, $B_{\phi}=0$. Also, $\lambda$ is the affine parameter, and the Christoffel symbols are given by the geometry defined in (\ref{metrika}). Here, according to the already discussed weak electromagnetic field approximation, the influence of the magnetic field on the metrics describing the Black Hole is neglected, and thus the only influence of the vacuum polarization comes through the change of the magnetic field distribution, which then also changes the motion of particles in such non-minimally coupled gravitational and magnetic field. Taking the proper time as the affine parameter and writing (\ref{lor}) in terms of the  four spacetime coordinates ($t$,$r$,$\phi$,$\theta$) = ($t(\tau)$,$r(\tau)$,$\theta(\tau)$,$\phi(\tau)$)), we obtain the following four equations:
\begin{widetext}
\begin{align}
&t''(\tau) = - \frac{2 M r'(\tau) t'(\tau)}{r(\tau)\left[r(\tau) - 2 M\right]}\,, \label{first} \\
\begin{split}
r''(\tau) =& \frac{M \left[r'(\tau)\right]^{2}}{r(\tau) \left[ r(\tau) - 2 M \right]} + \left[ r(\tau) - 2 M \right] \left\{ \left[\theta'(\tau)\right]^{2} + \left[ \phi'(\tau) \right]^{2} \sin^{2}\left[\theta(\tau)\right]  - \frac{M \left[t'(\tau)\right]^{2}}{r^{3}(\tau)}\right\} \\
&- \frac{q_{\rm s}\sqrt{1 - \frac{2 M}{r(\tau)}} \left[r^{3}(\tau) - 2 M q_{3} \right] B_{\rm rad}(r(\tau)) \phi'(\tau)}{2 r^{2}(\tau)} \sin^{2}\left[\theta(\tau)\right]\,,  \\
\end{split} \\
&\theta''(\tau) = \frac{\cos[\theta(\tau)] \sin[\theta(\tau)] \phi'(\tau) \left\{ q_{\rm s} \left[r^{3}(\tau) + 4 M q_{3} \right] B_{\rm rad}(r(\tau)) + r^{3}(\tau) \phi'(\tau) \right\}}{r^{3}(\tau)} - \frac{2 r'(\tau) \theta'(\tau)}{t(\tau)}\,, \\
\begin{split}
\phi''(\tau) = & - \frac{2 r'(\tau) \phi'(\tau)}{r(\tau)} - 2 \cot[\theta(\tau)] \theta'(\tau) \phi'(\tau)\\
&+ \frac{q_{\rm s} B_{\rm rad}(r(\tau)) \left\{ \left[ r^{3}(\tau) - 2 M q_{3} \right] r'(\tau) - 2 \sqrt{1 - \frac{2 M}{r(\tau)}} r(\tau) \left[r^{3}(\tau) + 4 M q_{3} \right]  \cot\left[ \theta(\tau) \right] \theta'(\tau) \right\}}{2 \sqrt{1 - \frac{2 M}{r(\tau)}} r^{4}(\tau)}\,,
\label{last}
\end{split}
\end{align}
\end{widetext}
where $\tau$ is the proper time and $q_{\rm s} = q/m$ is the specific electromagnetic charge of the particle.

\section{Numerical Simulations} \label{sec:NumSim}

\begin{figure*}[hbtp]
  \centering
  \includegraphics[scale=0.336]{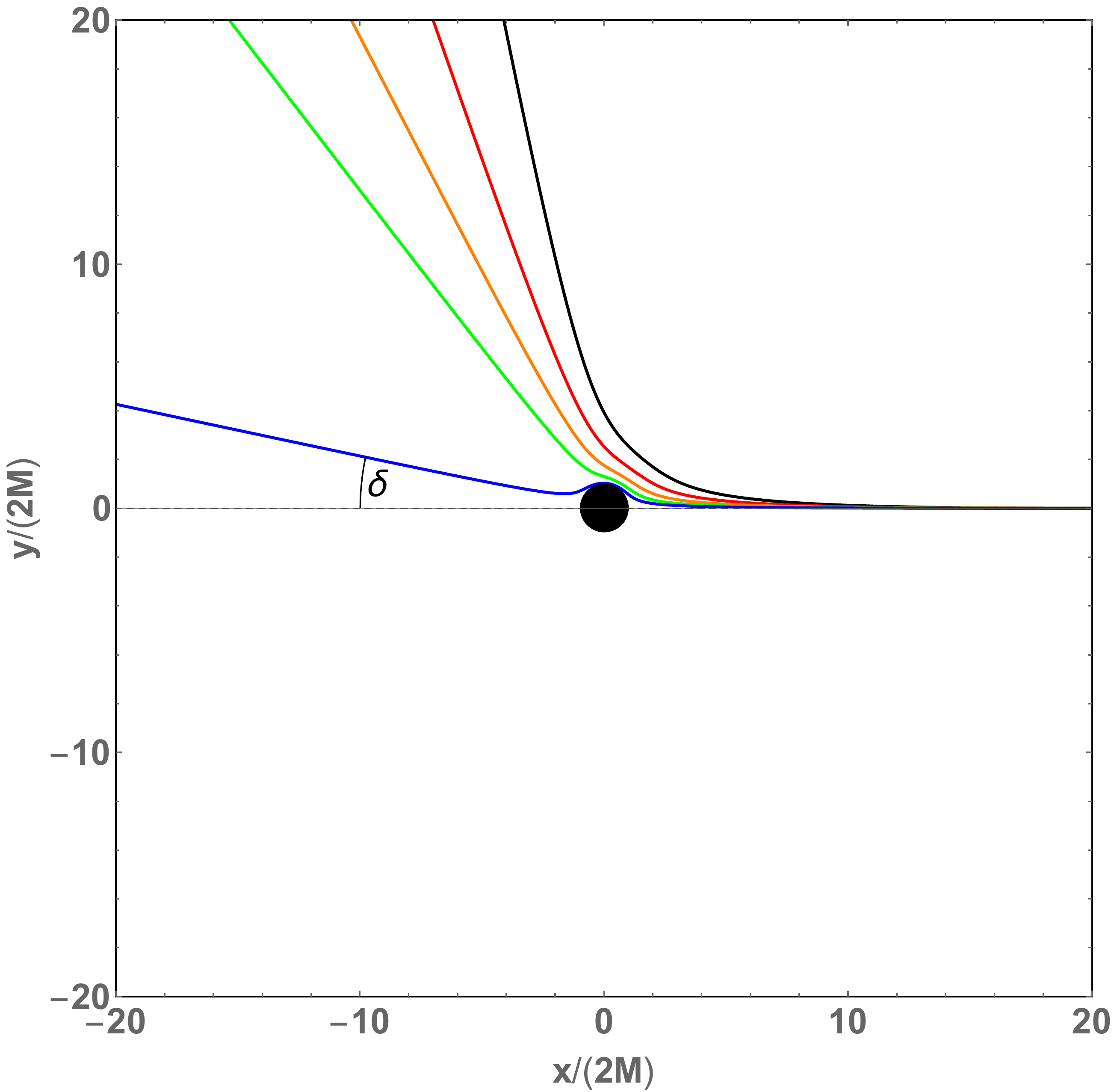}
  \includegraphics[scale=0.336]{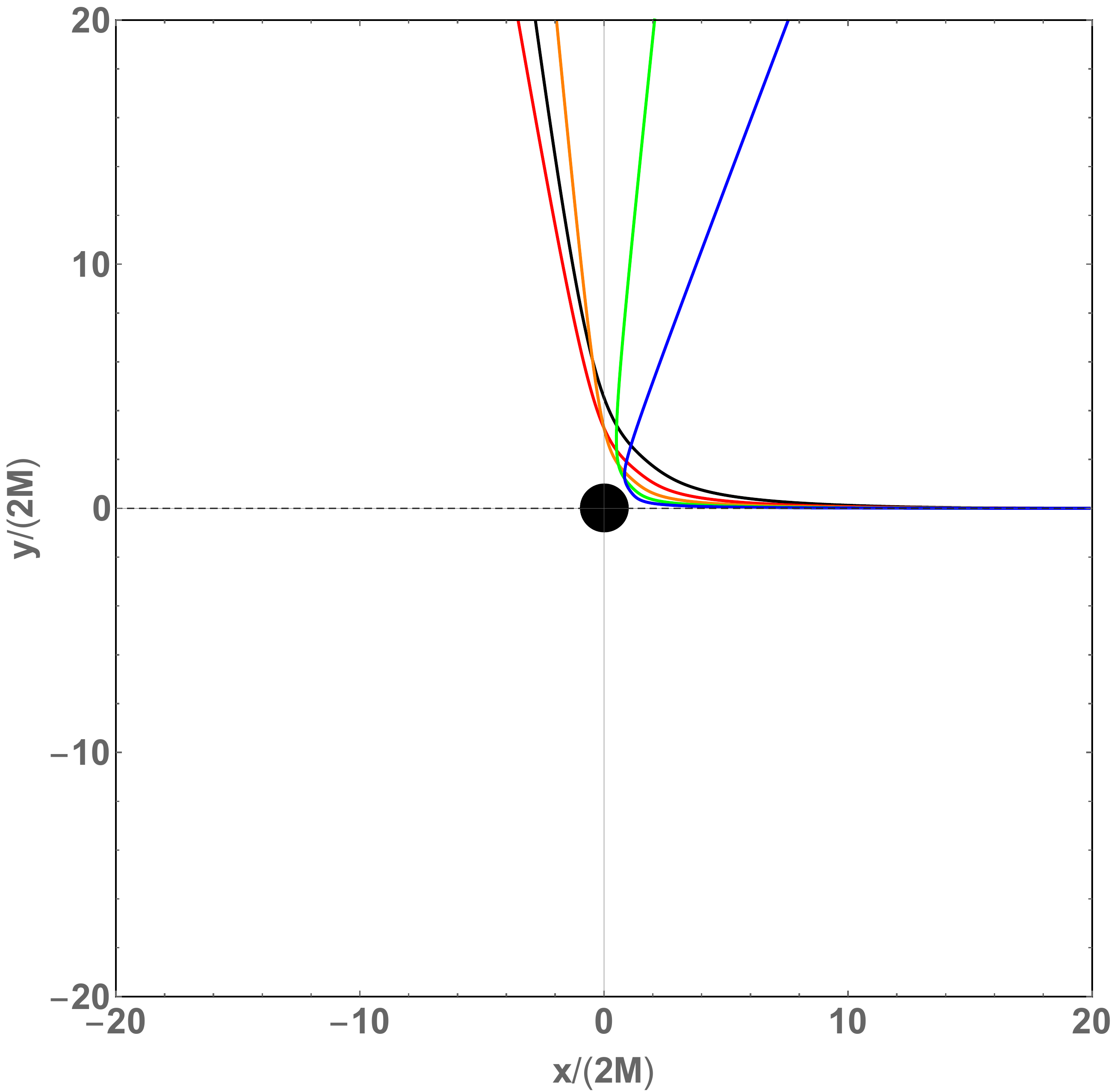}
  \includegraphics[scale=0.336]{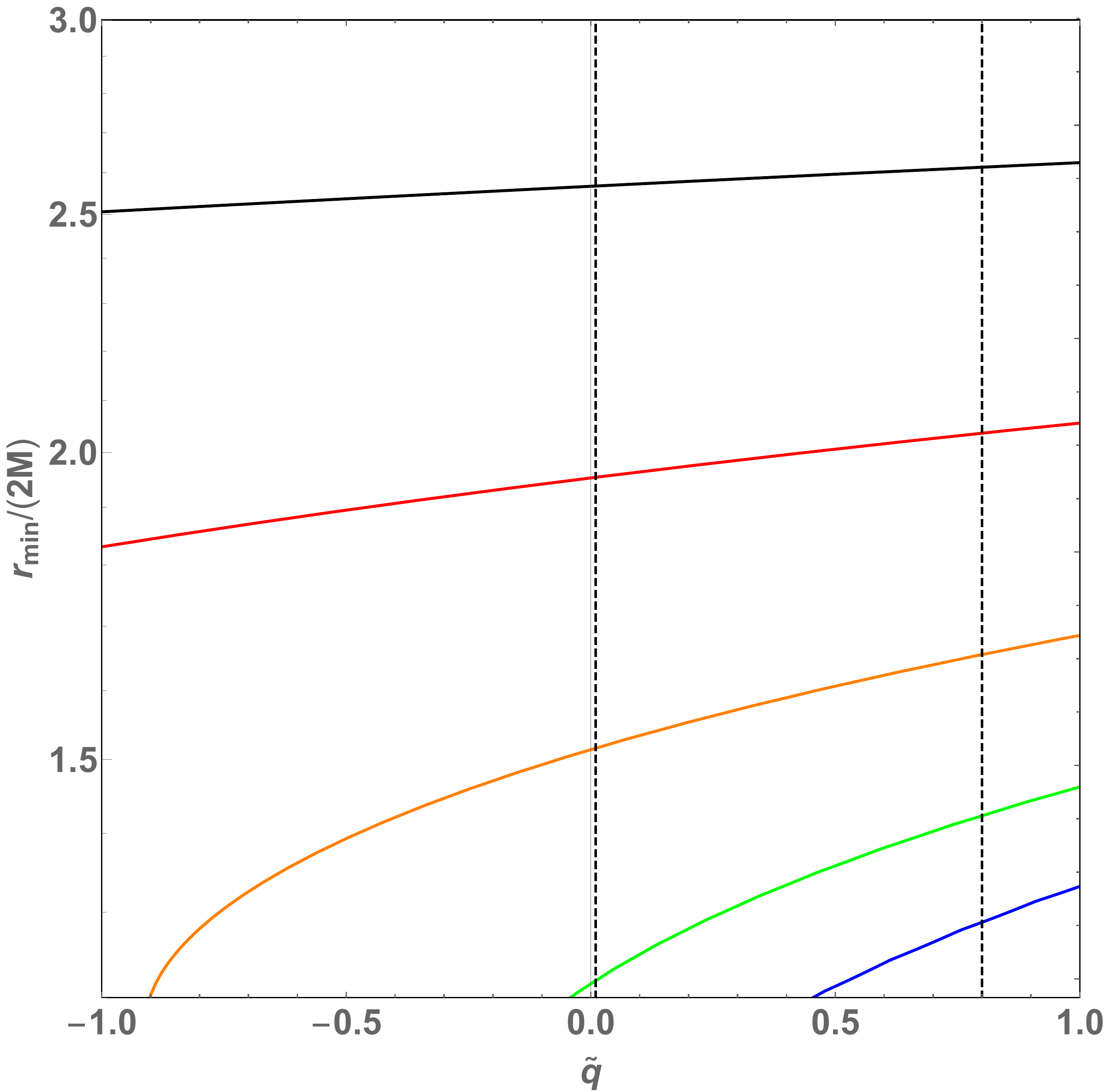}
  \includegraphics[scale=0.336]{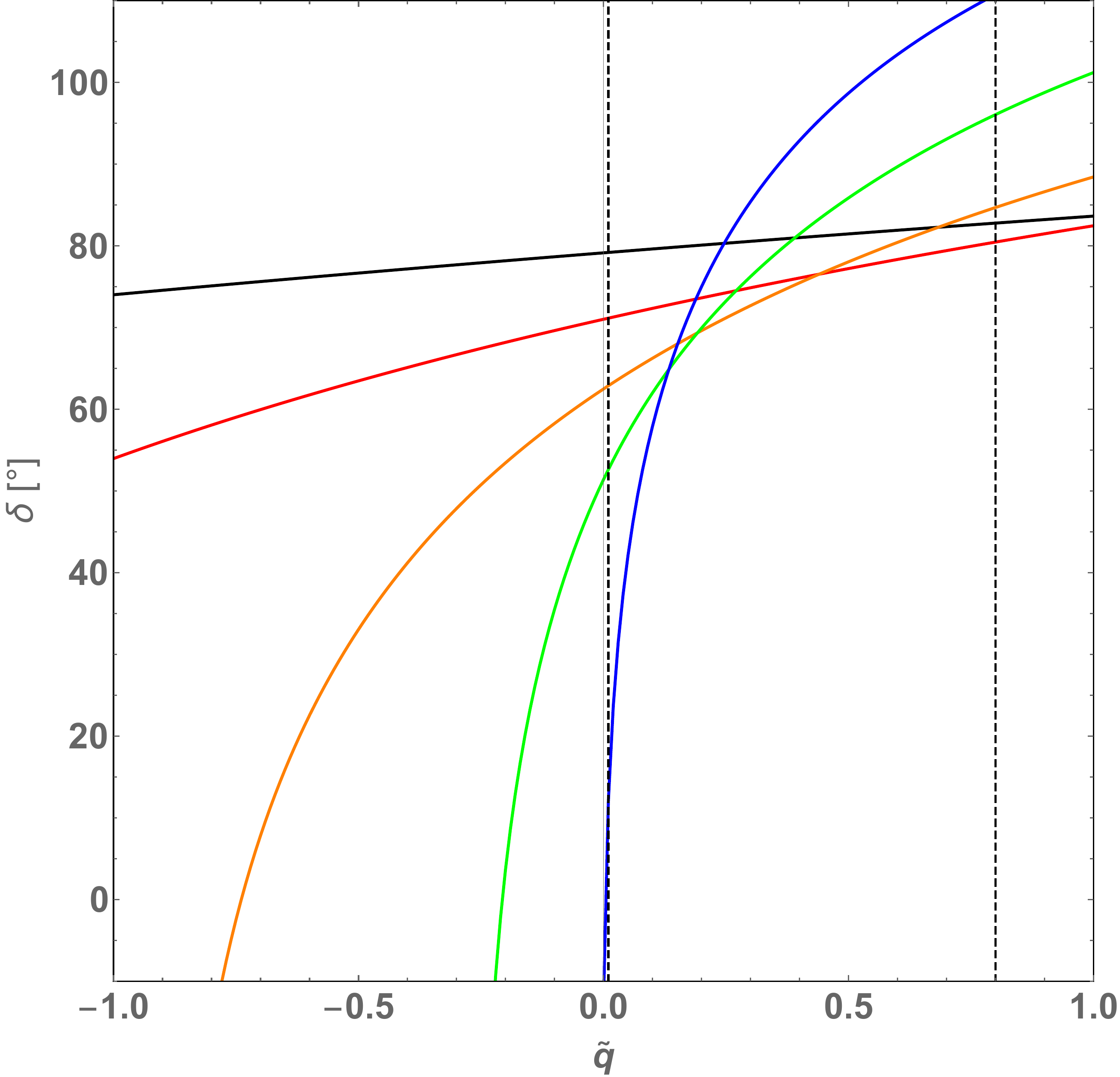}
  \includegraphics[scale=0.701]{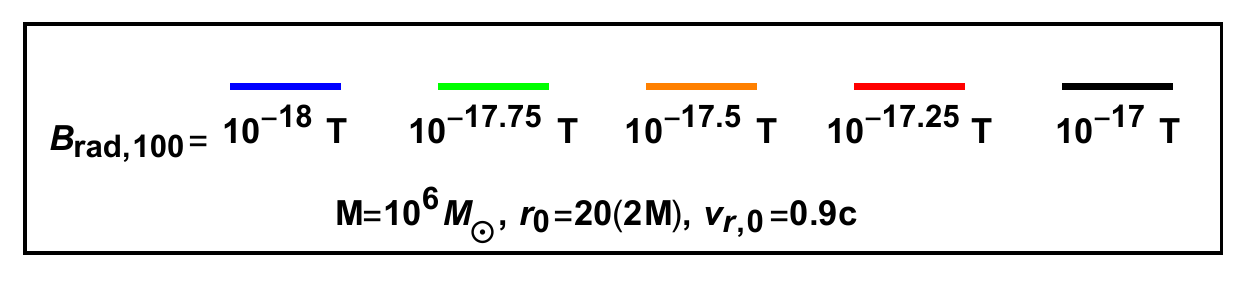}
  \includegraphics[scale=0.701]{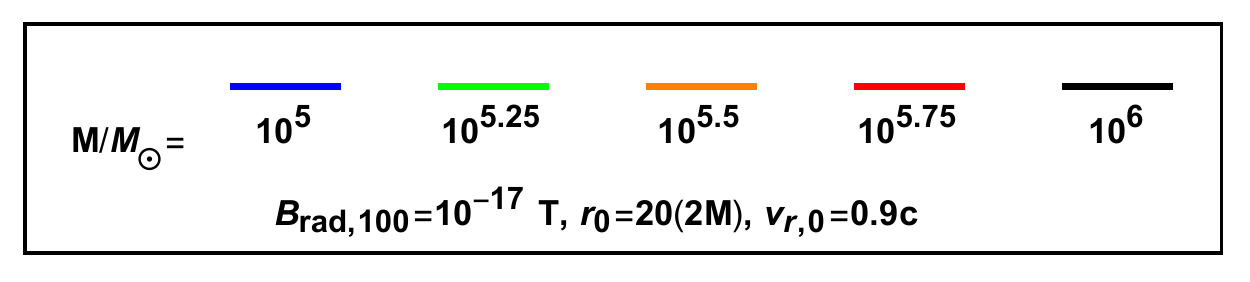}
  \caption{Results for trajectories of electrons for different magnetic field strengths/Black Hole masses. \emph{Top panel}: Sample trajectories for different magnetic field strengths with $M = 10^{6} M_{\odot}$ (left legend at the bottom)/different Black hole masses with $B_{\rm rad,0} = 10^{-17} \, \rm{T}$ (right legend at the bottom), at $\tilde{q} = 0.01$ (left) and $\tilde{q} = 0.8$ (right), where the former depicts how the deflection angle $\delta$ is measured. \emph{Bottom panel}: Minimum distance from the center of the Black Hole (left) and deflection angle (right), the vertical dashed lines indicate the values for $\tilde{q}$ used in the top panel. In all cases the default initial conditions $r_{0} = 20 \times (2 M)$ and $v_{r,0} = -0.9 c$ have been used. The black line indicates the fiducial scenario defined in Eq.~(\ref{fiducial_unbound}).}
  \label{fig:e_B}
\end{figure*}

By introducing the generalized four-velocity vector $(v_{t},v_{r},v_{\theta},v_{\phi})$ as well as using the differential equations (\ref{radijalna})-(\ref{thetajedn}) and (\ref{first})-(\ref{last}) for the magnetic field, we then get a set of ordinary differential equations for the description of the propagation of a charged particle in the presence of a SBH:
\begin{widetext}
\begin{align}
&t'(\tau) = v_{t}(\tau) \\
&v_{t}'(\tau) = - \frac{2 M v_{r}(\tau) v_{t}(\tau)}{r(\tau)\left[r(\tau) - 2 M\right]}\,, \\
&r'(\tau) = v_{r}(\tau) \\
\begin{split}
v_{r}'(\tau) =& \frac{M \left[v_{r}(\tau)\right]^{2}}{r(\tau) \left[ r(\tau) - 2 M \right]} + \left[ r(\tau) - 2 M \right] \left\{ v_{\theta}^{2}(\tau) +  v_{\phi}^{2}(\tau) \sin^{2}\left[\theta(\tau)\right]  - \frac{M v_{t}^{2}(\tau)}{r^{3}(\tau)}\right\} \\
&- \frac{q_{\rm s}\sqrt{1 - \frac{2 M}{r(\tau)}} \left[r^{3}(\tau) - 2 M q_{3} \right] B_{\rm rad}(r(\tau)) v_{\phi}(\tau)}{2 r^{2}(\tau)} \sin^{2}\left[\theta(\tau)\right]\,,  \\
\end{split}\\
&\theta'(\tau) = v_{\theta}(\tau)\,, \\
&v_{\theta}'(\tau) = \frac{\cos[\theta(\tau)] \sin[\theta(\tau)] v_{\phi}(\tau) \left\{ q_{\rm s} \left[r^{3}(\tau) + 4 M q_{3} \right] B_{\rm rad}(r(\tau)) + r^{3}(\tau) v_{\phi}(\tau) \right\}}{r^{3}(\tau)} - \frac{2 v_{r}(\tau) v_{\theta}(\tau)}{t(\tau)}\,, \\
&\phi'(\tau) = v_{\phi}(\tau)\,, \\
\begin{split} 
&v_{\phi}'(\tau) = - \frac{2 v_{r}(\tau) v_{\phi}(\tau)}{r(\tau)} - 2 \cot[\theta(\tau)] v_{\theta}(\tau) v_{\phi}(\tau)\,,\\
&+ \frac{q_{\rm s} B_{\rm rad}(r(\tau)) \left\{ \left[ r^{3}(\tau) - 2 M q_{3} \right] v_{r}(\tau) - 2 \sqrt{1 - \frac{2 M}{r(\tau)}} r(\tau) \left[r^{3}(\tau) + 4 M q_{3} \right]  \cot\left[ \theta(\tau) \right] v_{\theta}(\tau) \right\}}{2 \sqrt{1 - \frac{2 M}{r(\tau)}} r^{4}(\tau)}\,, 
\end{split} \\
&B_{\rm rad}'(r) = \frac{\left[ 10 M^{2} q_{3} - r^{4} + M \left( r^{3} - 4 q_{3} r \right) \right] + C \left[ 2 \sqrt{1 - \frac{2 M}{r}} r \left( 4 M q_{3} + r^{3} \right) \right]}{r \left(r - 2 M\right) \left( r^{3} - 2 M q_{3} \right)} B_{\rm rad}(r)\,. \label{Brad}
\end{align}
\end{widetext}
These differential equations require a total of nine initial/boundary conditions and a value for $C$ in ($\ref{Brad}$). For the following consideration we fix the values 
\begin{equation}
t_{0} \equiv t(\tau = 0) = 0\,, \, v_{t,0} \equiv v_{t}(\tau = 0) = 1\,, \, C = -1\,,
\end{equation}
while the values of the initial conditions for the other time-dependent quantities, as well as the condition for $B_{\rm rad}$, i.e.~the values for
\begin{equation}
\begin{split}
&r_{0} \equiv r(\tau = 0)\,, \, v_{r,0} \equiv v_{r}(\tau = 0)\,, \, \\
&\theta_{0} \equiv \theta(\tau = 0)\,, \, v_{\theta,0} \equiv v_{\theta}(\tau = 0)\,, \, \\
&\phi_{0} \equiv \phi(\tau = 0)\,, \, v_{\phi,0} \equiv v_{\phi}(\tau = 0)\,, \, \\
&B_{\rm rad,100} \equiv B_{\rm rad}(r = 100 \times (2 M))
\end{split}
\end{equation}
are considered as free parameters of the simulations. The boundary condition for $B_{\rm rad}$ has been chosen at $r = 200M$, i.e.~at 100 Schwarzschild Radii, as at that distance from the Black Hole the magnetic field is nearly a dipole, such that by fixing the magnetic field at that value we ensure the comparability with the case without the vacuum polarization effect. For the solution of the differential equations the method described in \cite{Petzold1983} has been used.

\begin{figure*}[hbtp]
  \centering
  \includegraphics[scale=0.336]{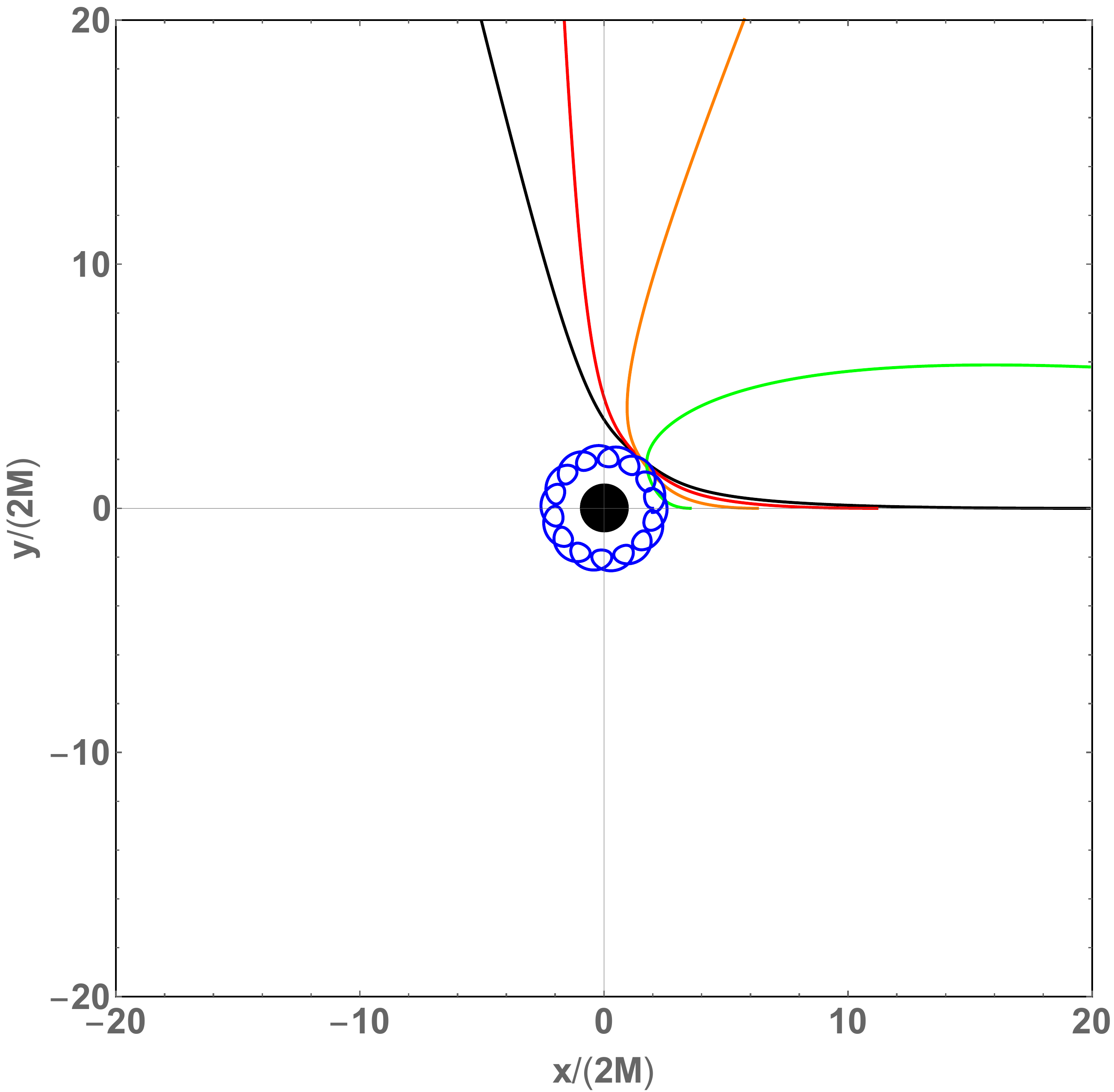}
  \includegraphics[scale=0.336]{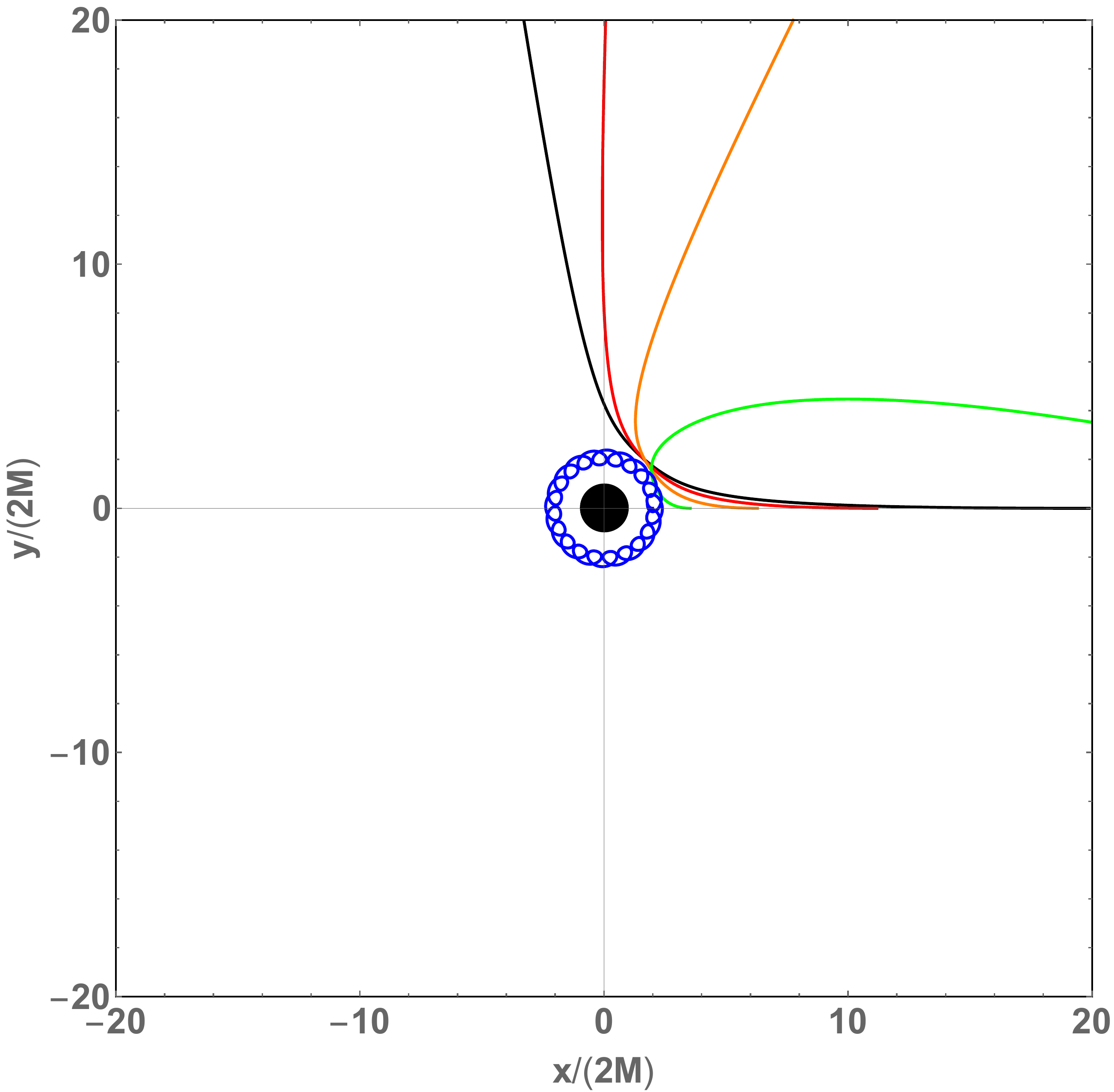}
  \includegraphics[scale=0.336]{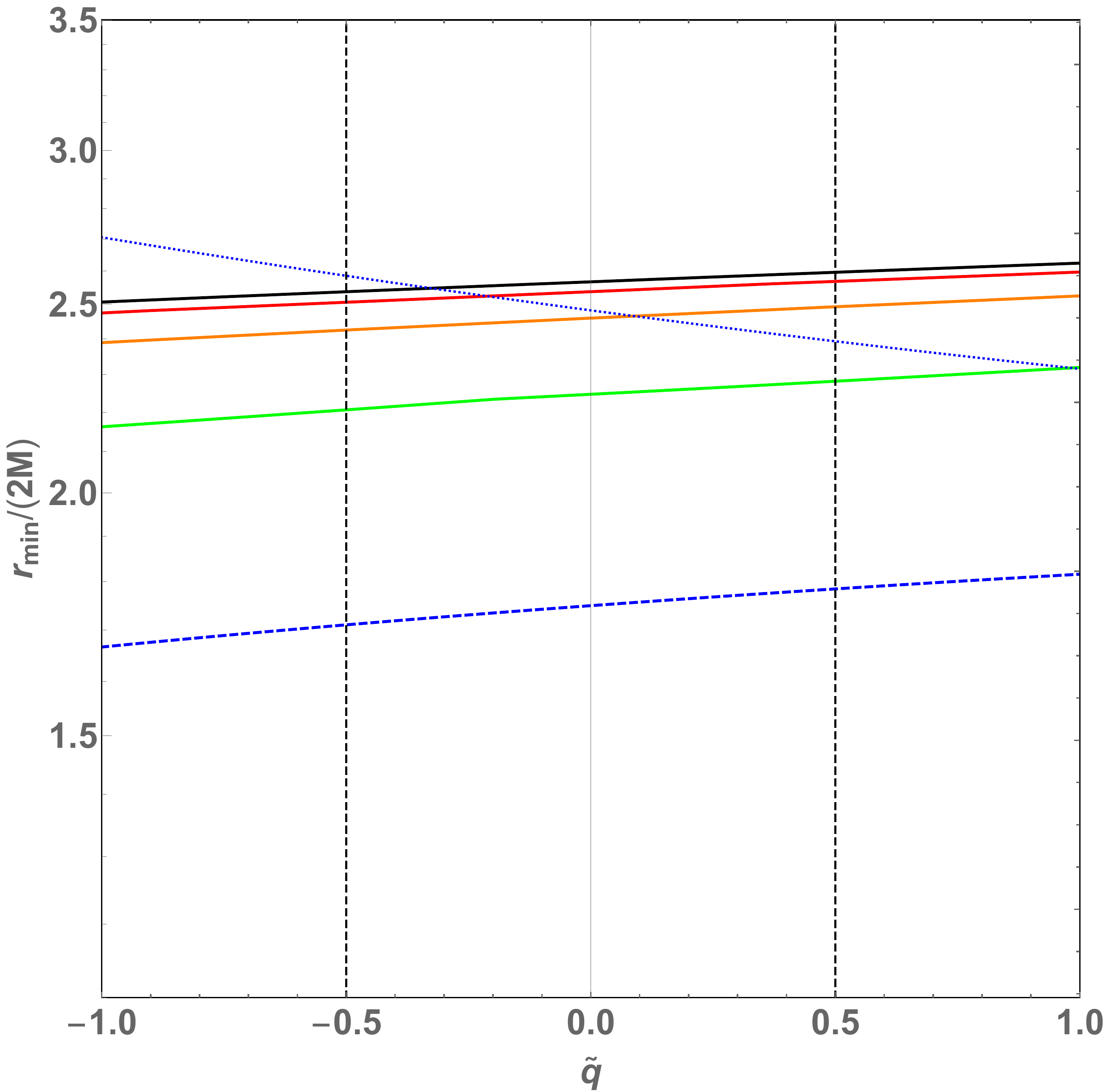}
  \includegraphics[scale=0.336]{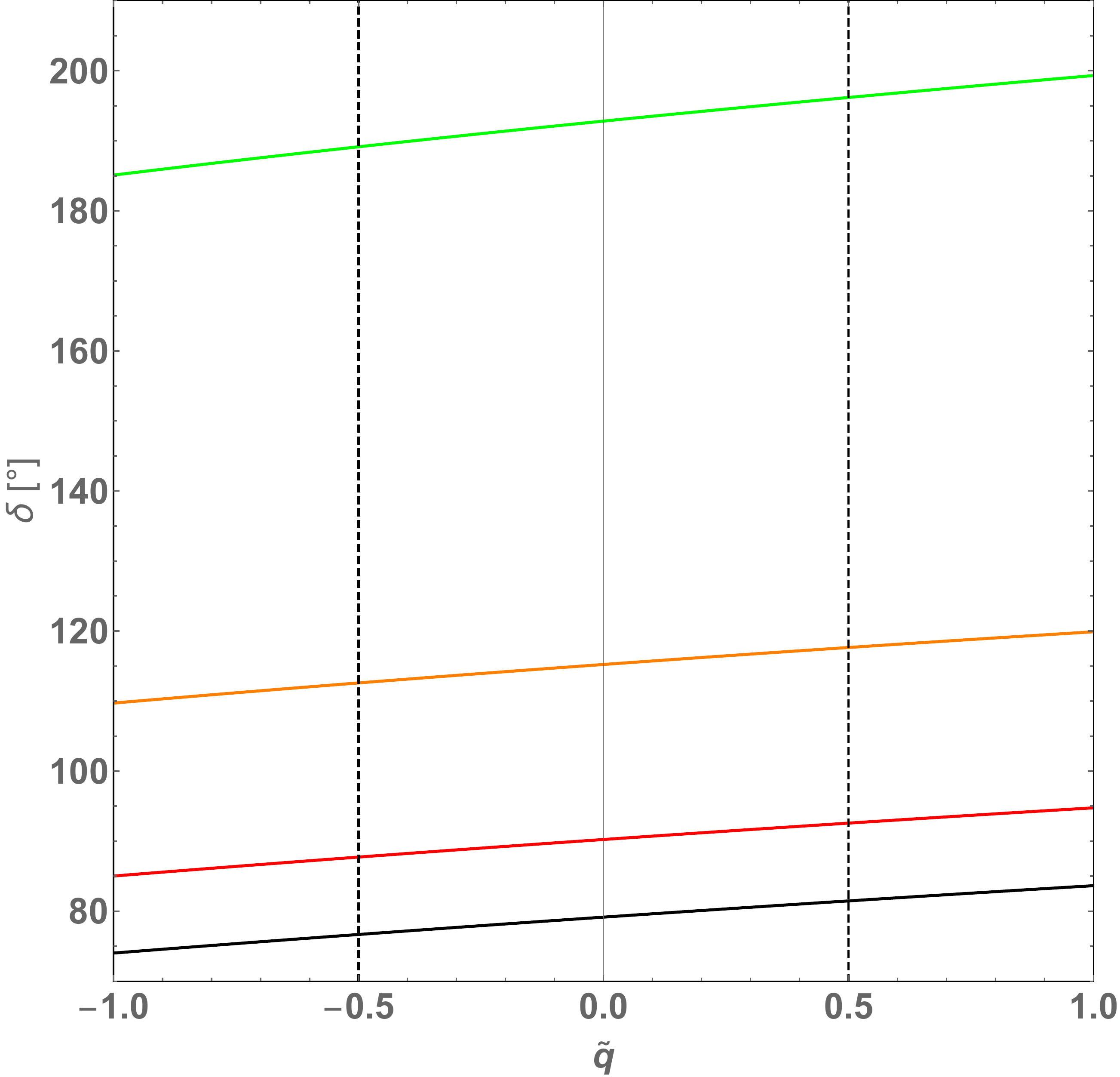}
  \includegraphics[scale=0.701]{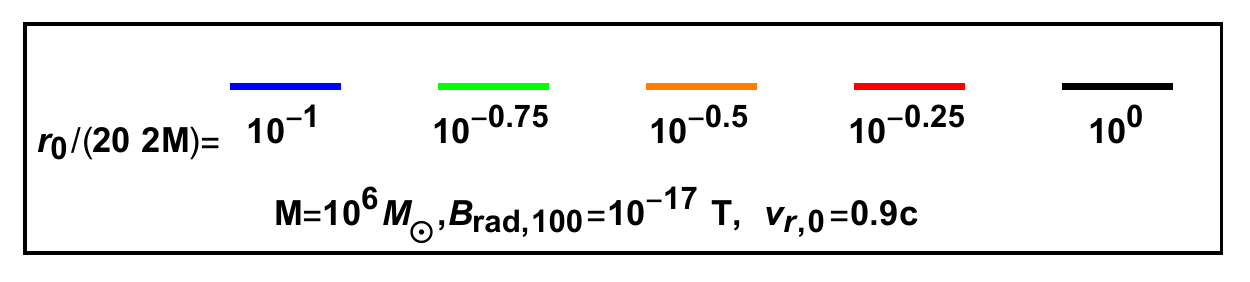}
  \caption{Results for trajectories of electrons for different initial distances from the Black Hole. \emph{Top panel}: Sample trajectories for different values of $r_{0}$, at $\tilde{q} = -0.5$ (left) and $\tilde{q} = 0.5$ (right). \emph{Bottom panel}: Minimum distance from the center of the Black Hole (left) and deflection angle (right), the vertical dashed lines indicate the values for $\tilde{q}$ used in the top panel. The black line indicates the fiducial scenario defined in Eq.~(\ref{fiducial_unbound}). The dashed blue line indicates that the trajectory is bound, such that in that case a maximum distance (blue dotted line) can be shown.}
  \label{fig:e_r}
\end{figure*}

Following \cite{Tursunov:2018erf} we limit the investigation to the cases where the particle propagates in the $xy$-plane (since trajectories outside that plane tend to become chaotic), such that for all the scenarios the additional restrictions
\begin{equation} \label{thetainit}
\theta_{0} = \pi/2\,, \, v_{\theta,0} = 0
\end{equation}
apply, leaving us with initial conditions $r_{0}$, $v_{r,0}$, $\phi_{0}$, $v_{\phi,0}$ and $B_{\rm rad,100}$. Using this, in the following we will investigate the two general cases of bound and unbound particle trajectories.

\begin{figure*}[hbtp]
  \centering
  \includegraphics[scale=0.336]{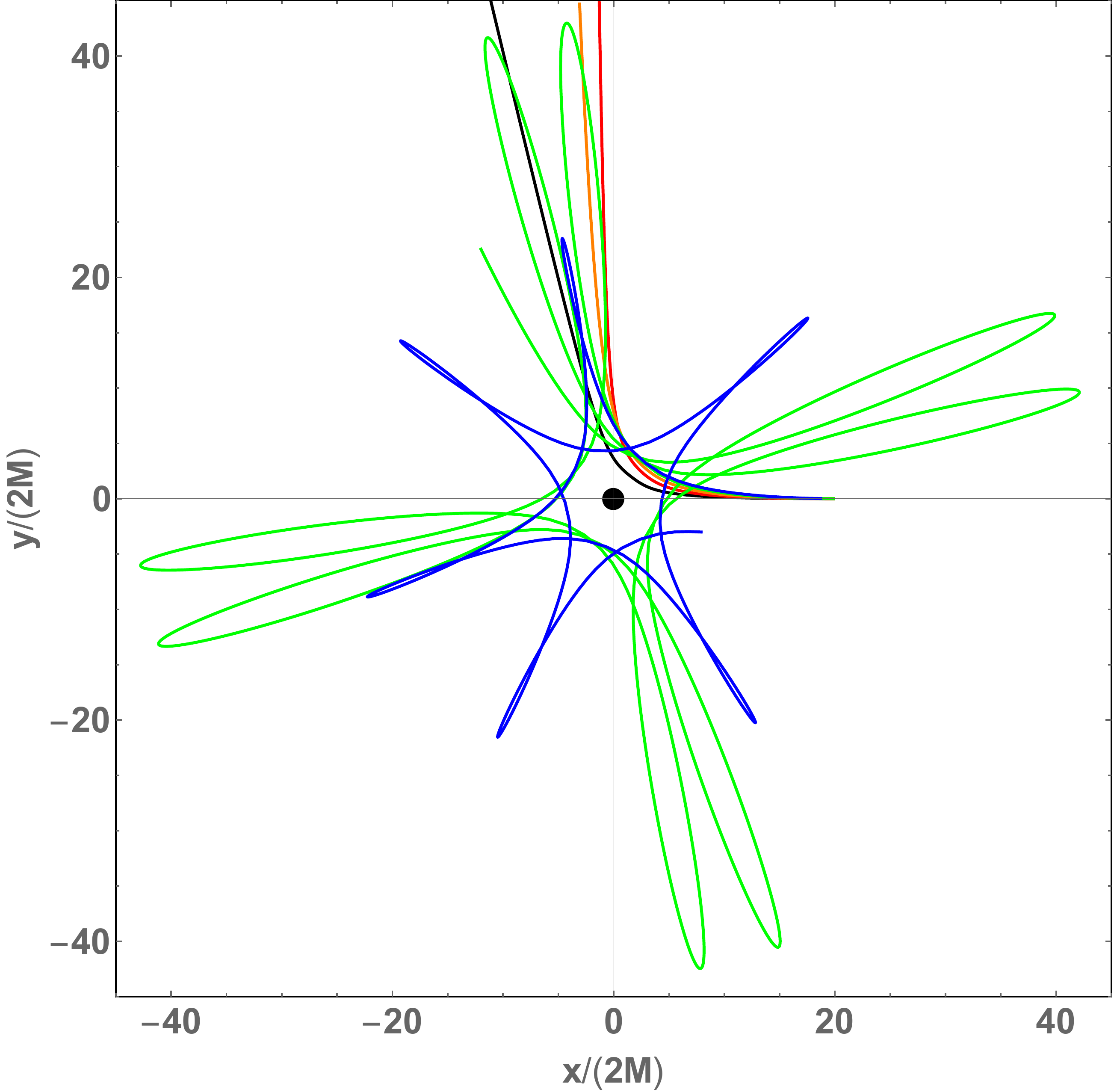}
  \includegraphics[scale=0.336]{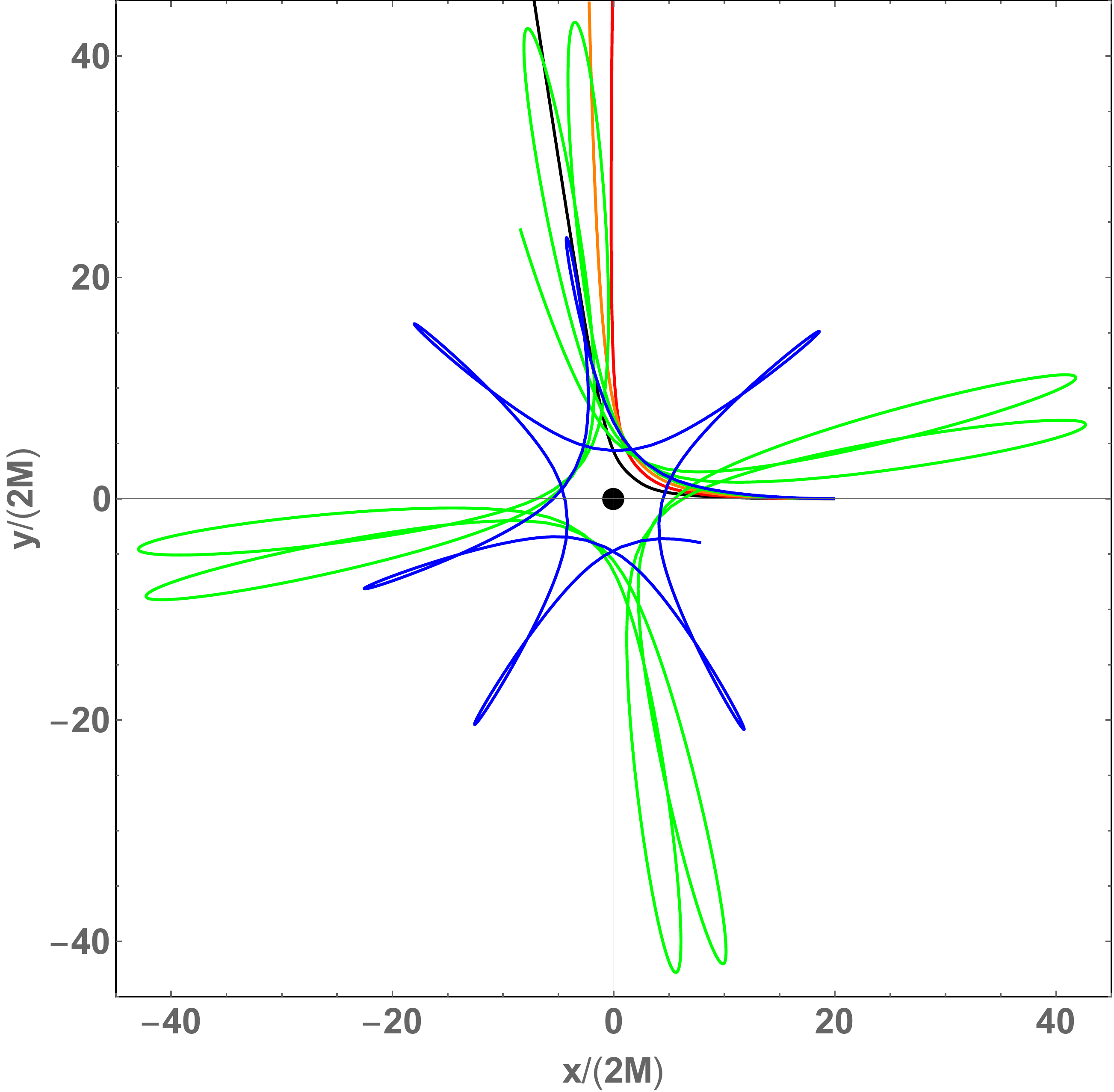}
  \includegraphics[scale=0.336]{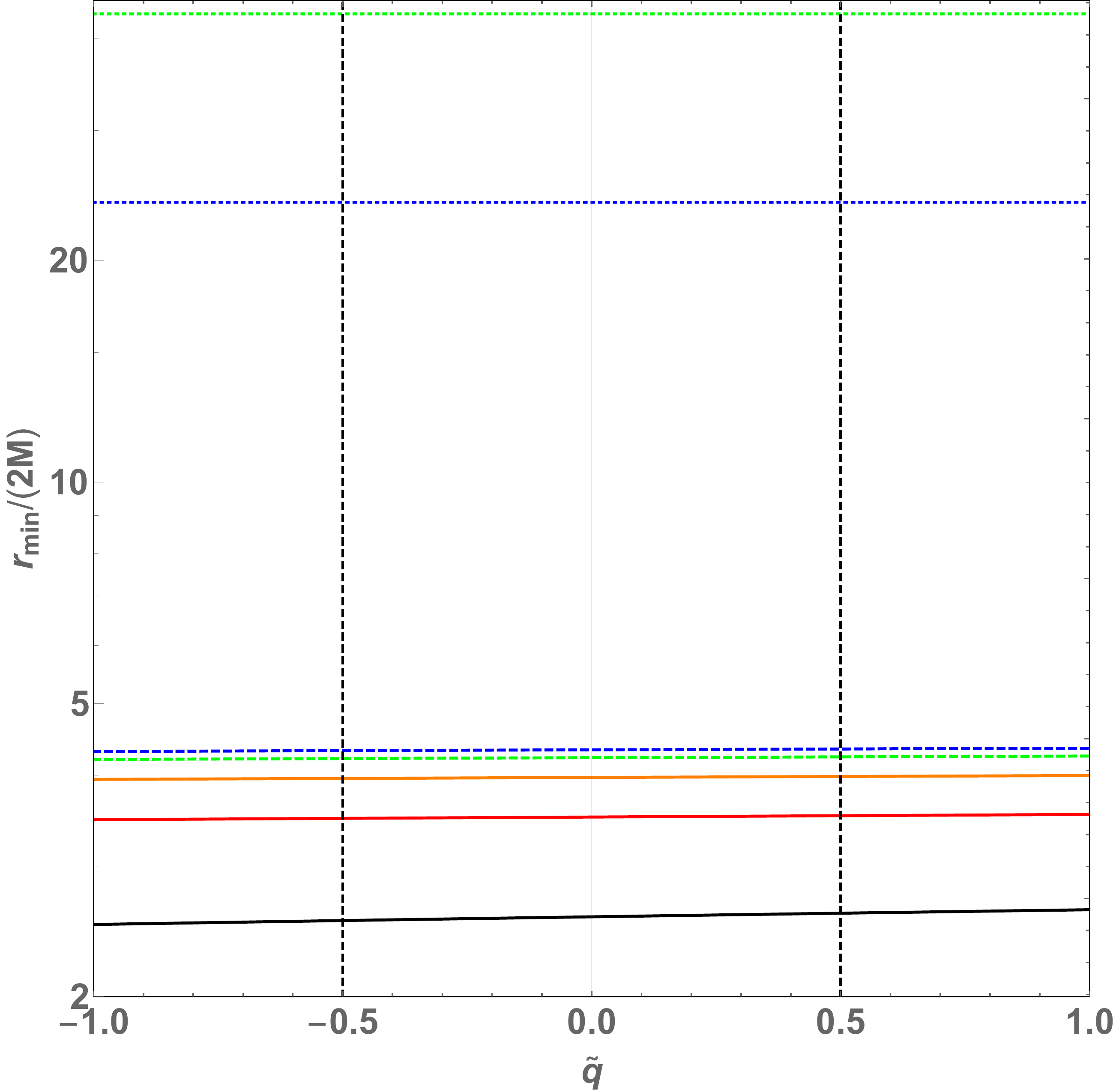}
  \includegraphics[scale=0.336]{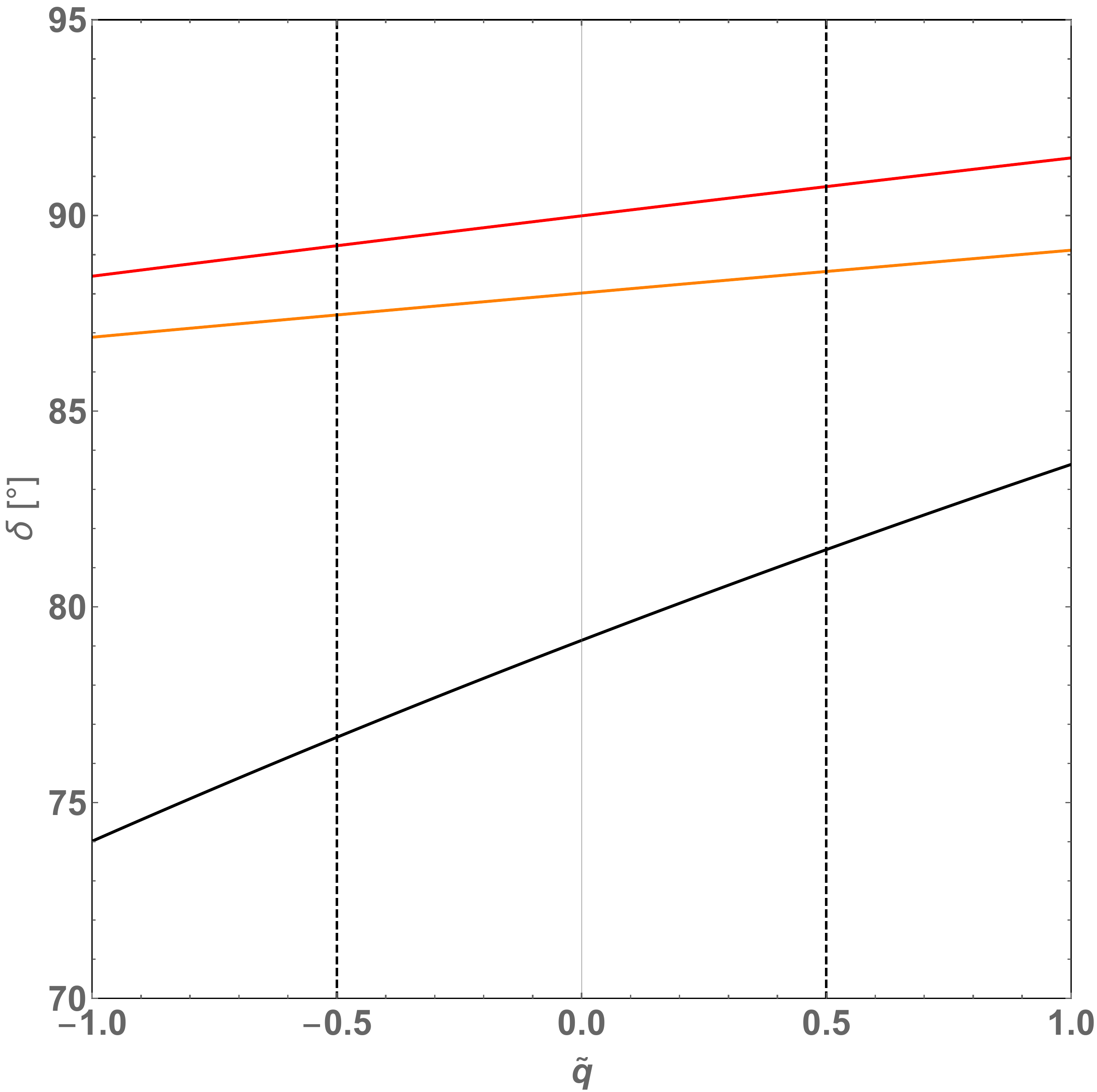}
  \includegraphics[scale=0.701]{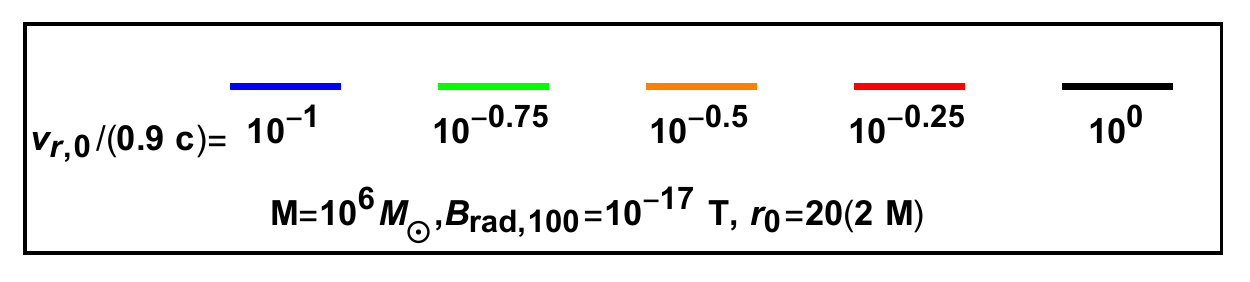}
  \caption{Results for trajectories of electrons for different initial velocities. \emph{Top panel}: Sample trajectories for different values of $v_{r,0}$, at $\tilde{q} = -0.5$ (left) and $\tilde{q} = 0.5$ (right). \emph{Bottom panel}: Minimum distance from the center of the Black Hole (left) and deflection angle (right), the vertical dashed lines indicate the values for $\tilde{q}$ used in the top panel. The black line indicates the fiducial scenario defined in Eq.~(\ref{fiducial_unbound}). The dashed lines indicates that the trajectory is bound, such that in that case a maximum distance (corresponding dotted line) can be shown.}
  \label{fig:e_v}
\end{figure*}

\section{Results} \label{sec:Results}

In this section we present the results of the numerical simulations of trajectories in the $xy$-plane (i.e.~fulfilling the initial conditions (\ref{thetainit})). To be more specific, we investigate the extreme case of the particle's velocity initially pointing towards the center of the Black Hole, while fixing $\phi_{0}$ to 0 without loss of generality (due to the symmetry of the magnetic field regarding $\phi$), which therefore gives the additional conditions
\begin{equation}
\phi_{0} = 0\,,\,v_{\phi,0} = 0\,,\,v_{r,0} < 0,
\end{equation}
leaving us with the two free propagation parameters, $r_{0}$ and $v_{r,0}$, together with the Black Hole mass $M$ and the magnetic field boundary condition $B_{\rm rad,100}$. In the following the fiducial values are given by
\begin{equation} \label{fiducial_unbound}
\begin{split}
M = 10^{6} M_{\odot}\,,\, r_{0} = 20 \times (2 M) \,, \\ 
v_{r,0} = -0.9 c \,,\, B_{\rm rad,100} = 10^{-17}\,{\rm T}\,, 
\end{split}
\end{equation}
where $M_{\odot}$is the solar mass.

\begin{figure*}[hbtp]
  \centering
  \includegraphics[scale=0.505]{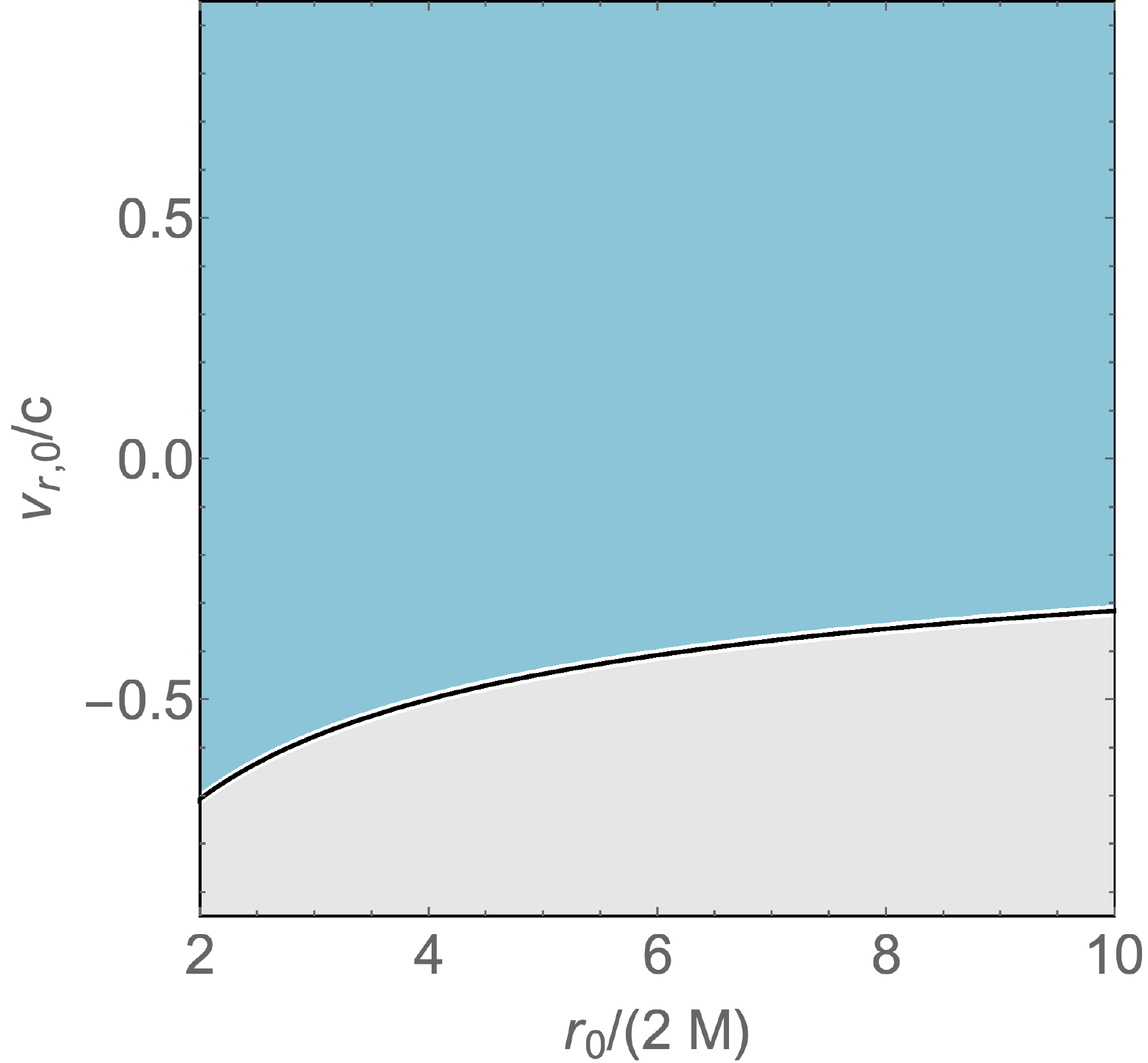}
  \includegraphics[scale=0.505]{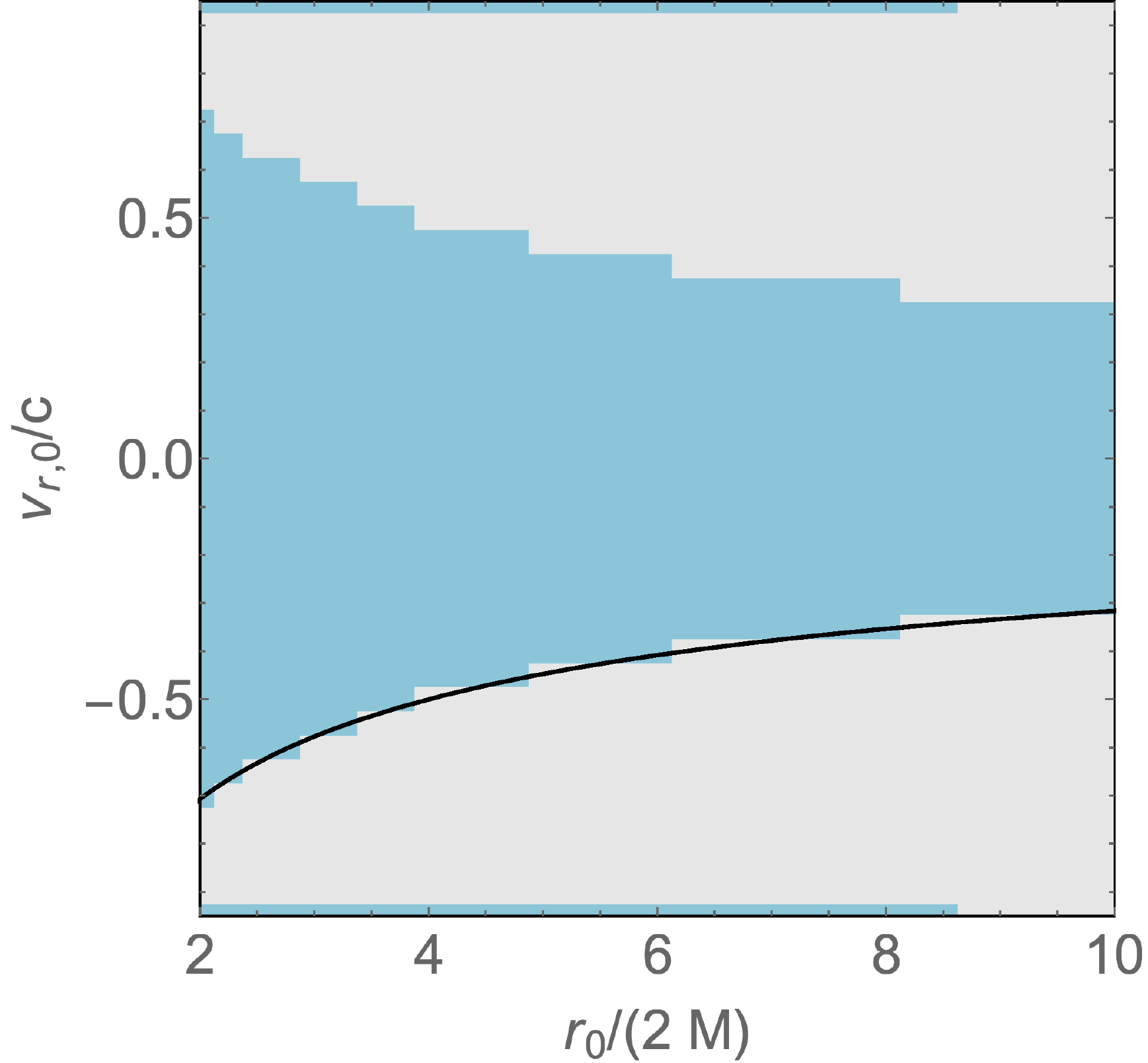}
  \includegraphics[scale=0.505]{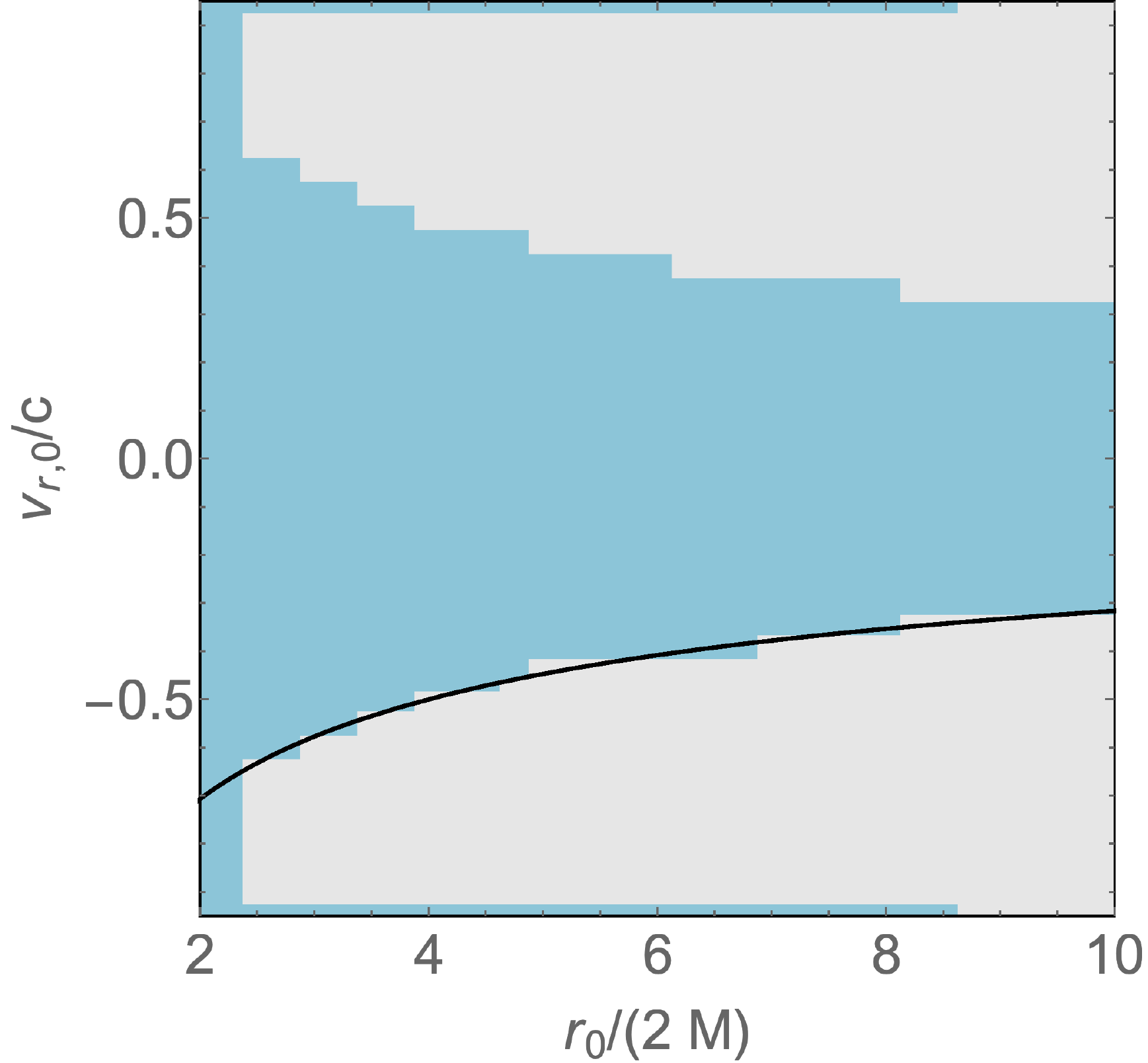}
  \includegraphics[scale=0.505]{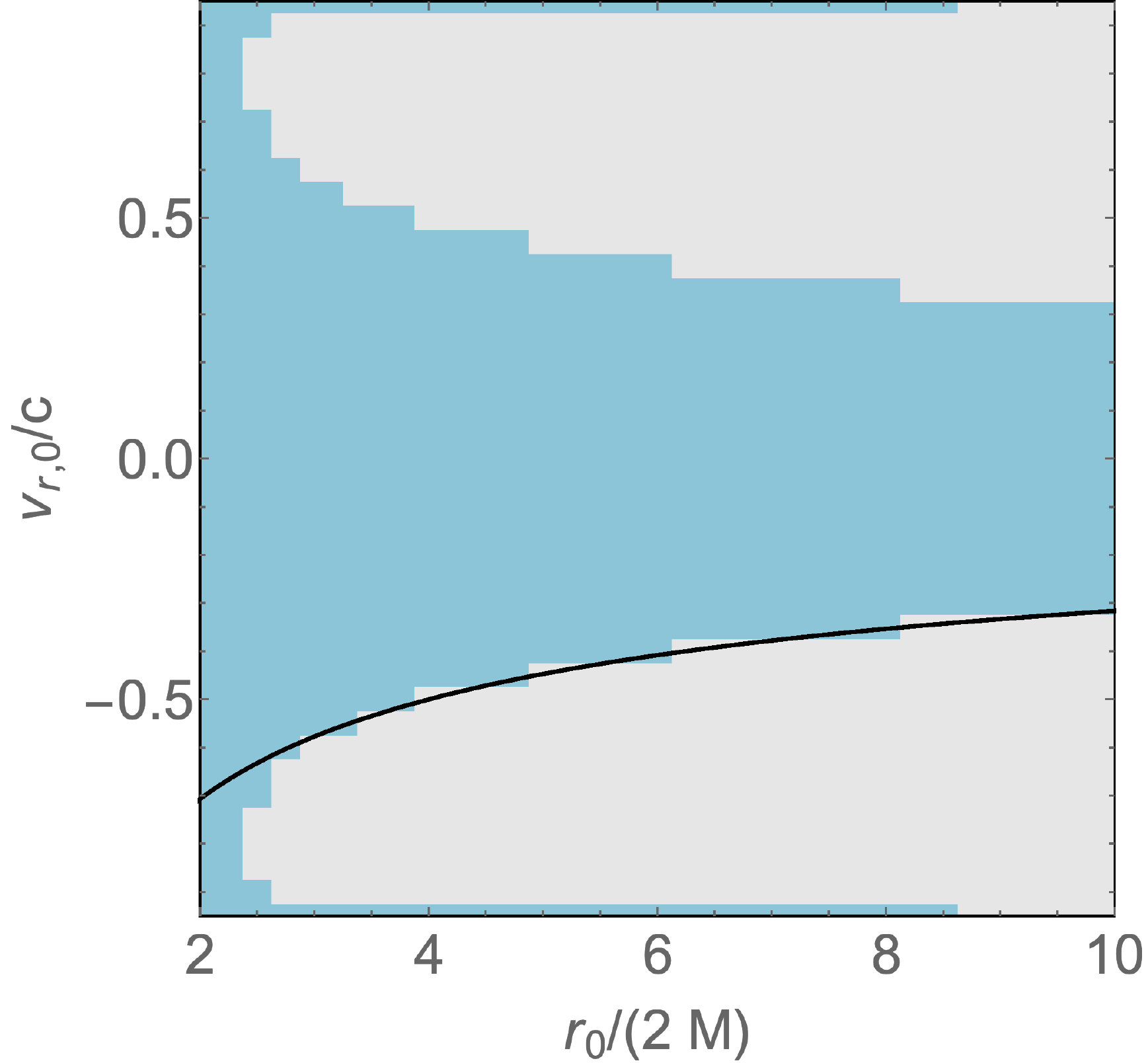}
  \caption{"Phase diagrams" for the condition of the trajectory of a charged particle at a SBH in the $r_{0}$ vs.~$v_{r,0}$ parameter space. Here, the gray region denotes an unbound state, i.e.~$r \rightarrow \infty$ for $\tau \rightarrow \infty$, while the turquoise region shows the parameter combination for which $r$ does \textit{not} go to infinity for $\tau \rightarrow \infty$, i.e.~either a bound trajectory around the Black Hole or the case where the particle falls into it. For the Black Hole the default values of $M = 10^{6} M_{\odot}$ and $B_{\rm rad,100} = 10^{-17}\,{\rm T}$ have been used. Note that $v_{r,0} < 0$ depicts the situation where initially the particle velocity points radially \textit{away} from the Black Hole, while for $v_{r,0} > 0$ it points \textit{towards} it. The thick black line in each subfigure shows the "phase transition" of the case without magnetic field. \textit{Top left}: The case without magnetic field. \textit{Top right}: The case for a dipole magnetic field. \textit{Bottom}: Phase diagram including the effect of vacuum polarization for $\tilde{q} = -1.0$ (\textit{left}) and $\tilde{q} = 1.0$ (\textit{right}). }
  \label{fig:PD}
\end{figure*}

As one can see from Fig.~\ref{fig:e_B}, the first result is that the scaling with the magnetic field strength is exactly the same as the scaling with the Black Hole mass $M$ as long as in both cases the initial position of the particle is the same measured in Schwarzschild Radii. These scalings are in fact the same as in the case without vacuum polarization, which can be best seen using the simple estimate for the influence of the two relevant forces here, namely, the Lorentz Force $F_{\rm L}$ and the gravitational force $F_{\rm G}$:
\begin{equation}
\frac{F_{\rm L}}{F_{\rm G}} \sim \frac{q_{\rm s} \gamma v r_{\rm 0}^{2}}{G} \frac{B(r_{0})}{M} \sim \frac{q_{\rm s} \gamma v r_{\rm s}^{2}}{G} \frac{B(r_{0})}{M} \sim M B(r_{0})\,.
\end{equation}
This ratio is proportional to the product $M B$ which explains the aforementioned scalings and shows that this basic principle remains valid.

Furthermore, from Fig.~\ref{fig:e_B} one can see that decreasing the value of $B_{\rm rad,100}$ (or the value of $M$) leads to a dramatic change of the trajectory -- once the magnetic field is low enough even small changes will highly impact the scattering angle. This can be understood looking at the lower left panel: Reducing the magnetic field results in a larger influence of the gravitational pull, such that the particle is able to get closer to the Black Hole, hence giving a smaller $r_{\rm min}$. Since, however, the magnetic field is significantly amplified at small $r$ and dramatically increases for $r\rightarrow 2M$, this results in a stronger deflection. In addition, this also explains why the impact of $\tilde{q}$ is much more visible for smaller $r_{\rm min}$ -- a larger value for $\tilde{q}$ causes a stronger amplification.  

In Fig.~\ref{fig:e_r} the trajectories of electrons for different initial positions $r_{0}$ are shown. While for values $r_{0} \gtrsim 4 \times (2 M)$ the expected behavior, i.e.~scattering at a moderate angle, is seen, for values of $r_{0}$ below that the trajectory is first bent up until the point where it reverses its initial direction ($\delta = 180^{\circ}$), or even beyond that, while for $r_{0} \leq 2 \times (2 M)$ in all analyzed cases it results in a bound state. This is a major difference compared to the classical dipole field as in that case the particle would still be only deflected or, with $r \sim 2M$, fall into the Black Hole. In the case of vacuum polarization on the other hand the magnetic field close to the Event Horizon grows strongly, therefore resulting in a large Lorentz Force. This Lorentz Force then remains in a delicate interplay with gravitation -- the latter being responsible for the circular movement around the Black Hole, while the former causes the small curls along the trajectory, as seen in the top panel of Fig.~\ref{fig:e_r}. As discussed above, the increase of $\tilde{q}$ results in a stronger amplification of the magnetic field, leading to a reduction of the Larmor Radius
\begin{equation}
r_{\rm L} = \frac{p}{q B}
\end{equation}
and therefore a decrease in the size of the loops caused by the Lorentz Force.

Finally, in Fig~\ref{fig:e_v} the influence of the initial velocity $v_{r,0}$ of the particle on its trajectory is presented. The first important result, as seen from the bottom left panel, is that for a fixed value of $v_{r,0}$ the value of $\tilde{q}$ does not play a significant role for $r_{\rm min}$. This means that at least for the given set of parameters the velocity penetrates the magnetic field only up to a point where the amplification is not fully dominant. Still, as for the other changes of parameters, from the bottom right panel one can see that for smaller $r_{\rm min}$ the impact of $\tilde{q}$ on the deflection angle $\delta$ is more significant. Furthermore, one can see that while for larger values of $v_{r,0}$ we obtain an unbound state, for small values we again have a bound one. Here, the situation is to some degree similar to the one described for the variation of $r_{0}$ -- for small values of $v_{r,0}$ the kinetic energy of the particle is reduced to a degree where the total energy is too small to escape the attraction potential of the magnetized Black Hole (while for small values of $r_{0}$ the total energy is too small due to the reduced amount of potential energy.

A different way to present and analyze the obtained results is to consider the bound (i.e.~encircling the Black Hole or falling into it) and unbound trajectories as "phases" in the phase space of the initial particle state described by the parameters $r_{0}$ and $v_{r,0}$. Then we can summarize the possible outcomes in a "phase diagram" where the change from one regime to the other may be understood as "phase transitions". This is done for different cases (for the fiducial values $M = 10^{6} M_{\odot}$ and $B_{\rm rad,100} = 10^{-17}\,{\rm T}$) in Fig.~\ref{fig:PD}.

The case without magnetic field (top left) can be roughly estimated analytically since in this case it is determined only by gravitation and hence the transition happens when the initial velocity corresponds to the escape velocity for $r = r_{0}$, i.e. in the Newtonian approximation

\begin{equation}
v_{r,0} =  - v_{\rm esc}(r_{0}) = - \sqrt{\frac{2 M}{r_{0}}} c\,.
\end{equation}

When including a dipole magnetic field (top right), two major changes compared to the non-magnetic case can be seen: First, the phase diagram obtains a symmetric shape regarding $v_{r,0}$, i.e. whether for a given $r_{0}$ the particle will end up in a bound or unbound state depends only on the magnitude of the initial velocity, but not on its sign. This is an important result as it means that a magnetic field which is strong enough may deflect the particle, even if it is initially directed towards the Black Hole, by such a large angle that, since it has sufficient kinetic energy, it can escape the gravitational potential. On the other hand, and that is the second major change, for large values of $\left| v_{r,0} \right|$ a new type of bound trajectory may emerge due to the large Lorentz Force which results in an enormous bending of the trajectory up to the point where it cannot escape. Due to the aforementioned symmetry this occurs even if for $v_{r,0} < 0$, as, due to its dominating magnitude, it is irrelevant in which direction the deflection occurs.

Once vacuum polarization is included, the situation is changed in particular for small $r_{0}$ (bottom panel) -- the minimum $r_{0}$ for which the particle can escape the gravitational pull of the Black Hole significantly increases, as mentioned above, due to the fact that the strength of the magnetic field is big enough to keep the particle bound due to the Lorentz Force. Looking at the phase diagram, one can in addition see that for some values of $r_{0}$, even as high as $r_{0} \gtrsim 2.0 \times (2 M)$, the particle ends up in a state which is not unbound independent of the initial velocity. Also, once can see that the value of $\tilde{q}$ has an influence on the setting -- increasing its value also increases the minimal $r_{0}$ for escaping.

\section{Conclusions and Outlook} \label{sec:Conclusions}
In this work we have studied the motion of charged particles in the non-minimally coupled gravitational and magnetic field that can result as a consequence of the vacuum polarization effect. We have assumed 
that magnetic fields are small enough so that the backreaction on the spacetime of the SBH -- which is the spacetime considered in this work - can be ignored. This is a completely justified assumption for the case of astrophysical magnetic fields. In this case, the effect of non-minimal coupling between gravity and magnetic fields will manifest through the change of the magnetic field, leading to its amplification or suppression, depending on the sign of the non-minimal coupling parameter. This will then, in turn, influence the motion of charged particles on the considered spacetime. We have furthermore chosen such a configuration of the magnetic field which for large distances from the Black Hole horizon approaches the physically important case of dipole magnetic field. Deriving the equations of motion describing this case, we performed the corresponding numerical simulations for physically realistic settings. The main conclusion of our work is that the non-minimal coupling can, for appropriate values of the coupling parameter, lead to significant changes in the motion of charged particles around the SBH. Since Black Holes surrounded by magnetic fields seem to represent the most promising setting for the manifestation of this effect, due to the corresponding high curvatures of spacetime, such results could be used to search for the signature of the vacuum polarization effect in combined magnetic and gravitational fields or to constrain the possible values of the coupling parameter by observations. \\ \\ 
The results of our numerical simulations have shown that the presence of non-minimal coupling between the magnetic field and gravity causes the change in trajectories of the charged particles around the Black Hole, as well as in deflection angle and minimal distance from the center of the Black Hole. The influence of this effect is more dramatically pronounced 
for smaller values of the magnetic field defined at some larger distance from the Black Hole center and smaller values of the Black Hole's mass. 
This can be explained by the fact that the change induced by the non-minimal coupling is strongest near the event horizon. Decreasing the magnetic field strength results in a larger influence of gravitational pull, and enables the particle to reach the region around the event horizon where the modification of the magnetic field is more pronounced.
Furthermore, it was demonstrated that for small enough initial distances from the center of the Black Hole, $r_{0} \leq 2 \times (2M)$, all analyzed cases result in a bound state, which is in strong contrast with the standard minimally coupled dipole, where only some of those trajectories would be deflected. 

Finally, we have systematically analyzed the configurations of initial 
position and velocity leading to bounded and unbounded orbits. It was shown that the vacuum polarization effect leads to changes in the type of trajectories, specially for the small initial distances from the Black Hole, leading to emergence of bounded states for the configurations 
of the parameters for which in the minimally coupled case only unbounded trajectories exist. 

The fact that for a given initial velocity the value of $\tilde{q}$ seemingly does not play such a big role as for the other parameters might be relevant for future investigations when an ensemble of particles with a given velocity distribution is considered.

It has also been found that, for otherwise the same initial conditions, whether the trajectory of a particle is bound or unbound might depend solely on $\tilde{q}$, which could be used for deriving constraints on $\tilde{q}$ itself. This might become even more significant once energy losses are taken into account. 

It has been shown that for some set of parameters charged particles cannot escape the magnetized Black Hole independent of their initial velocity. In particular, this happens for particles starting off at a distance below a characteristic value determined by the magnetic field and the Black Hole mass. This means that the observed flux of particles emitted close to the Black Hole (or their secondaries) is expected to be suppressed compared to the unmagnetized case.

In the future, we are planning to consider other magnetic field configurations, such as the asymptotically homogeneous magnetic field which has been studied in previous publications. Another important extension of the work presented here is the inclusion of energy loss mechanisms due to acceleration of the charged particles as this would produce synchrotron radiation which may be observable at Earth. In this context we are also going to analyze a distribution of particle velocities (instead of a single particle as in this work) in order to obtain more realistic resulting radiation signatures which might be compared with actual observations. Finally, in order to take into account more complicated processes like accretion, actual general-relativistic magnetohydrodynamic simulations are necessary. 

The value of the coupling parameter between gravity and electromagnetic fields should be considered as a free parameter, and if its actual value is very small, the discussed effects will not be empirically obvious. However, even in this most conservative option, which would put the value of the coupling constant only to the order of the Compton wavelength for electrons (see discussion in \cite{Drummond:1979pp} and \cite{Pavlovic:2018idi}), the effect can become strongly pronounced in the case of primordial black holes, characterized by very small masses and sizes, surrounded by primordial magnetic fields. Therefore, this scenario deserves to be investigated in future, with an emphasis on its cosmological implications. 

\begin{acknowledgments}
The work of AS was supported by the Russian Science Foundation under grant no.~19-71-10018, carried out at the Immanuel Kant Baltic Federal University.

The authors would like to thank Sumarna Haroon for discussions and interest in this work.
\end{acknowledgments}

\end{document}